\documentclass[english]{article}
\usepackage{lmodern}
\usepackage[]{fontenc}
\usepackage[latin9]{inputenc}
\usepackage[letterpaper]{geometry}
\usepackage{amstext}
\usepackage{graphicx}
\usepackage{color}

\makeatletter

\providecommand{\tabularnewline}{\\}

\usepackage{slashed}

\usepackage{ulem}

\makeatother

\usepackage{babel}

\begin{document}

\title{The possible bound state of the double heavy meson-baryon system }
\author{Qing Xu$^1$, Hong-ying Jin$^1$ and T. G.  Steele$^2$\\
$^1$Zhejiang Institute of Modern Physics, Zhejiang University, Zhejiang Province, P. R. China\\
$^2$ Department of Physics and Engineering Physics,
University of Saskatchewan,\\
Saskatoon, Saskatchewan, Canada S7N 5E2 }
 \maketitle
\begin{abstract}
We calculate the  two-pion exchange potential between a heavy meson
and a heavy baryon. We find this potential is as strong as the
one-pion exchange potential between two heavy mesons and is enough to
bind $\Lambda_b-\bar B$. Though our result is
sensitive to  the cut-off, the value of the cut-off is in the reasonable
region.

\end{abstract}

\section{Introduction}
In the naive quark model, there are no more than three quarks and anti-quarks in a hadron.
This picture is not quite consistent
with the fundamental theory of the strong interaction, Quantum
Chromodynamics (QCD), because colour singlets can be formed from larger numbers of quark constituents.
Thus it is generally believed that
hadron configurations should exist  beyond the naive quark model,
i.e., hadrons  composed of four, five or more quarks and
anti-quarks. Actually,  the nuclei, which can be considered as
multi-quark states, have already been found long ago. It is amazing
that the four and five quark states have not been found yet. It is possible that QCD dynamics prevents the formation of these states or their experimental observation may be difficult. For QCD itself, the confirmation of
such ``exotic" hadrons in experiments  surely  is welcome .

The simplest extension of the naive quark model is the four-quark
state  \cite{Fist raised}. Some  four-quark states may have
``exotic" quantum numbers and can be easily distinguished from the
``normal" mesons, but because most of them have the same quantum number as
the ``normal" mesons, identification of them becomes challenging. At
present, there is a focus on the so-called  molecule states, which have an
anomalously large decay width in a special channel in which the
masses of the states are on the kinematic threshold. The famous
examples are $f_0(980)$ and $X(3872)$. Such  molecule states are
considered as loose bound states of two normal mesons via pion
exchange  \cite{Systematicly studied}. This picture is very similar to
the nuclei.

 For the five-quark state  \cite{pentaquark}, the train of thought is similar.  Some states
 [e.g.,  the candidate  $\theta(1540)$]
  do not mix with the normal baryons and can be identified easily. But most of them
 [e.g., the candidate  $N^*(1535)$)]
  may be confused with the normal baryons. The configuration of the five quark
 states has two types, the diquark correlation and the meson-baryon bound state  system. The $N^*(1535)$ is
 often considered as the bound state of $K\Sigma$  \cite{1535}. However,  most  discussions  on meson-baryon molecules is  restricted to light meson-baryon systems. This could be attributed to either experimental difficulties or challenging dynamics.  In the meson-meson system, the one-pion exchange plays the most important role {in binding two mesons. In the meson-baryon system,
 since there is no one-pion exchange (OPE), the four-particle interaction, such as the  $\bar NN\bar K K$ interaction in the case of the  $\Lambda(1405)$ (which is considered as a quasi-bound state of $\bar K N$ \cite{1405}), is very important.  The coupling of the four-particle interaction can be fixed  within Chiral Perturbation Theory (CPT) only if the constituent meson of  the  system is a goldstone boson.  For the double heavy meson-baryon system, such as the $\Lambda_c \bar D $ system which we will consider,  the  four-particle interaction is currently unknown and its
   simple  extension  may have a large uncertainty. Instead, we consider two-pion exchange (TPE) in the
   $\Lambda_c \bar D $ system.

   TPE  not only can  provide intermediate and long distance interactions, but also can provide the short distance interaction. The magnitude of
   TPE may be referred  to the nucleon's case. In nuclei,  TPE  is important partly  because of  the large coupling $g_{N\Delta\pi}$ \cite{wise1} \cite{backman}. For  our case, the coupling between  heavy baryons  is also very large  \cite{Lagrangian of Lambda and Sigma}. As a naive dimensional analysis  \cite{kolck2},
   the ratio between TPE of  $D\Lambda_c$ and OPE of $D\bar D^*$ is $g_4^2M_{\sigma}\mu/(8\pi f_\pi^2)$
   (we use OPE for $D\bar D^*$  because there is no OPE between $\Lambda_c D$). The energy scale $\mu^2v= (M_\Sigma-M_\Lambda)^2-M_\pi^2\approx (200 MeV)^2$,
   $g_4\approx=1.24\pm 0.17$, so $D\bar D^*$ is $g_4^2M_{\sigma}\mu/(8\pi f_\pi^2) \sim 1$. For  $D\bar D^*$, many authors claim \cite{Close:2003sg}\cite{Thomas&Close}\cite{Systematicly studied}\cite{Tornqvist:2004qy} that OPE may be strong enough to bind  $D\bar D^*$ together.  So it is appropriate to consider  effects of TPE between $D\Lambda_c$.  Without any information about  counter terms (the  four-particle interaction), our calculation is  clearly  dependent on the cutoff. A reasonable cutoff may be chosen by referring to the deuteron and X(3872) cases.

   From the experimental point, SELX has already claimed the existence of the double heavy baryon $\Xi_{cc}$  \cite{mattson}. Along with more and more double heavy baryons  being  found in experiments, the situation may become very similar to  heavy quarkonium, i.e., many resonances may be considered as molecule states, such as X(3872), $Z_b(10608)$ \cite{BB}. Even the double heavy baryon-baryon state has been already considered \cite{ll}.  The possibility of the double  heavy meson-baryon state
   will  be considered eventually.

    In this paper, we consider the bound state of the $\bar D(\bar B)-\Lambda_c(\Lambda_b)$ system via TPE
    between the $\bar D(\bar B)$ and the $\Lambda_c(\Lambda_b)$.  The two-pion exchange potential (TPEP) is regularized by  a Gaussian
    type form factor  \cite{Ordonez:1995rz} \cite{T.A.Rijken}.  The potential is
    sensitive to the cut-off $\Lambda$ as expected, but we find it is  as strong as
    the one-pion exchange potential (OPEP) of the $\bar D-D^*$ and  $\bar B-B^*$ system at $\Lambda\sim 1GeV$, and the latter is believed strong enough to bind the two heavy mesons. Then we discuss the bound state of of the $\bar D(\bar
    B)-\Lambda_c(\Lambda_b)$ system by solving the Sch$\ddot{o}$dinger
   equation and find a bound state for  $\Lambda\sim
   1GeV$. Finally, we give a brief discussion and conclusion.

\section{Two-pion exchange potential and the bound state of $\bar D-\Lambda_c$}
The effective chiral lagrangian for the heavy mesons and baryons
were already given in  \cite{M.B.Wise} \cite{Lagrangian of Lambda and
Sigma}.  For the heavy meson system, the lagrangian can be
systematically expanded in the powers of small external momenta
\begin{equation}\label{meson}
\mathcal{L}=-i\text{Tr}\overline{H}_{a}\nu_{\mu}\partial^{\mu}H_{a}
+\frac{1}{2}i\text{Tr}\overline{H}_{a}H_{b}\nu^{\mu}(\xi^{\dagger}\partial_{\mu}\xi
+\xi\partial\xi^{\dagger})_{ba}+\frac{1}{2}ig\text{Tr}\overline{H}_{a}H_{b}\gamma_{\nu}\gamma_{5}(\xi^{\dagger}\partial_{\mu}\xi-\xi\partial\xi^{\dagger})_{ba}
+\cdots
\end{equation}
with \[
H_{a}=\frac{1+\slashed{\nu}}{2}(P_{a\mu}^{*}\gamma^{\mu}-P_{a}\gamma_{5})\]
 \[
\overline{H}_{a}=(P_{a\mu}^{*\dagger}\gamma^{\mu}+P_{a}^{\dagger}\gamma_{5})\frac{1+\slashed{\nu}}{2},\]
where
\[
M=\left[\begin{array}{ccc}
\frac{\pi^{0}}{\sqrt{2}}+\frac{\eta}{\sqrt{6}} & \pi^{+} & K^{+}\\
\pi^{-} & -\frac{\pi^{0}}{\sqrt{2}}+\frac{\eta}{\sqrt{6}} & K^{0}\\
K^{-} & \overline{K^{0}} &
-\sqrt{\frac{2}{3}}\eta\end{array}\right],\]
$\Sigma=e^{2iM/f_{\pi}}$ with $f_{\pi}=132\, MeV$ and
$\xi=\Sigma^{1/2}$. $\nu_\mu$ is the velocity of the heavy meson, $g$ is the coupling constant and
\[
V_{\mu}=\frac{1}{2}(\xi^{\dagger}\partial_{\mu}\xi+\xi\partial_{\mu}\xi^{\dagger}),\]

\[
A_{\mu}=\frac{i}{2}(\xi^{\dagger}\partial_{\mu}\xi-\xi\partial_{\mu}\xi^{\dagger}).\]

 In this paper, we only consider the leading order of the heavy meson expansion   and   the order of  the chiral expansion up to  $O(p^2/(4\pi f_\pi)^2)$,  so after substituting  D meson fields for $P$ and $P^{*}$, we can write  the interaction  part of  Eq.(\ref{meson})
 in the  rest reference frame of the heavy hadron (which is also approximately   the center of mass reference frame of system we will discuss)
 as
\begin{eqnarray*}
\mathcal{L}_{DD*} & = & \frac{-2}{f_\pi^{2}}\overline{D}t\cdot(\pi\times\dot{\pi})D-\frac{2g}{f_\pi}(\sqrt{2}D^{*\dagger}t\cdot\vec{\nabla}\pi D+\sqrt{2}D^{\dagger}D^{*}t\cdot\vec{\nabla}\pi)-\frac{2}{f_\pi^{2}}\overline{D}^{*}t\cdot(\pi\times\dot{\pi})D^{*},\end{eqnarray*}
  where $2t$ is Pauli matrix.

For the heavy baryon, the lagrangian is written as \cite{Lagrangian of Lambda and Sigma}
\begin{equation}\label{baryon}
\begin{array}{lll}
\mathcal{L} & = & \frac{1}{2}\text{Tr}[\overline{B}_{\overline{3}}(i\slashed{D}-M_{\overline{3}})B_{\overline{3}}]+\text{Tr}[\overline{B}_{6}(i\slashed{D}-M_{6})B_{6}]\\
 &  & +\text{Tr}\{\overline{B}_{6}^{*\mu}[-g_{\mu\nu}(i\slashed{D}-M_{6}^{*})+i(\gamma_{\mu}D_{\nu}+\gamma_{\nu}D_{\mu})-\gamma_{\mu}(i\slashed{D}+M_{6}^{*})\gamma_{\nu}]B_{6}^{*\nu}\}\\
 &  & +g_{1}\text{Tr}(\overline{B}_{6}\gamma_{\mu}\gamma_{5}A^{\mu}B_{6})+g_{2}\text{Tr}(\overline{B}_{6}\gamma_{\mu}\gamma_{5}A^{\mu}B_{\overline{3}})+H.C.+g_{3}\text{Tr}(\overline{B}_{6\mu}^{*}A^{\mu}B_{6})+H.C.\\
 &  & +g_{4}\text{Tr}(\overline{B}_{6}^{*\mu}A_{\mu}B_{\overline{3}})+H.C.+g_{5}\text{Tr}(\overline{B}_{6}^{*\nu}\gamma_{\mu}\gamma_{5}A^{\mu}B_{6\nu}^{*})+g_{6}\text{Tr}(\overline{B}_{\overline{3}}\gamma_{\mu}\gamma_{5}A^{\mu}B_{\overline{3}}),\end{array}\end{equation}
where $g_i, i=1,6$ are coupling constants, $B_3$, $B_6$ and $B^*_6$  are the fields of anti-triplet , sextet baryons with 1/2 spin and sextet baryons with 3/2 spin respectively. Explicitly,
\[
B_{6}=\left[\begin{array}{ccc}
\Sigma_{Q}^{+} & \frac{1}{\sqrt{2}}\Sigma_{Q}^{0} & \frac{1}{\sqrt{2}}\Xi_{Q}^{'+1/2}\\
\frac{1}{\sqrt{2}}\Sigma_{Q}^{0} & \Sigma_{Q}^{-} & \frac{1}{\sqrt{2}}\Xi_{Q}^{'-1/2}\\
\frac{1}{\sqrt{2}}\Xi_{Q}^{'+1/2} &
\frac{1}{\sqrt{2}}\Xi_{Q}^{'-1/2} & \Omega_{Q}\end{array}\right],\]
\[
B_{\overline{3}}=\left[\begin{array}{ccc}
0 & \Lambda_{Q} & \Xi_{Q}^{+1/2}\\
-\Lambda_{Q} & 0 & \Xi_{Q}^{-1/2}\\
-\Xi_{Q}^{+1/2} & -\Xi_{Q}^{-1/2} & 0\end{array}\right],\]
$B^*_6$ is similar to $B_6$. $M_3$ , $M_6$ and $M^*_6$   are the masses of anti-triplet , sextet baryons with 1/2 spin and sextet baryons with 3/2 spin respectively.
Similarly, keeping the leading order of the chiral expansion, we obtain the interaction part of Eq.(\ref{baryon})
\[
\mathcal{L}=\frac{-g_{2}}{f_\pi}\Sigma_{c}\gamma_{\mu}\gamma_{5}\partial^{\mu}\pi\Lambda_{c}+H.C.+\frac{-g_{4}}{f_\pi}\Sigma_{c}^{*}\partial^{\mu}\pi\Lambda_{c}+H.C.+\frac{-1}{f_\pi^{2}}\Lambda_{c}t\cdot(\pi\times\dot{\pi})\Lambda_{c}.\]

Normally, the coupling constants $g$ and $g_i(i=1,6)$ should be determined by the experimental data. From the decay width $\Gamma(D^*\to D\pi)$ \cite{data}, one can obtain $g=0.59\pm0.06$. (In the following, we only use the central value $g=0.59$. This value is widely used, for instance see  \cite{kolck2}.)  In the absence of the experimental data, $g_i(i=1,6)$ could be determined by the heavy quark symmetry and quark model \cite{Lagrangian of Lambda and Sigma}. From  \cite{Lagrangian of Lambda and Sigma},  $g_{4}=-\sqrt{3}g_{2}$, $g_4=1.24\pm 0.17$. $g_1$, $g_3$, $g_5$ and $g_6$ are not needed in this paper.

Since $\Lambda_c$ is an isospin singlet, there is no $\Lambda_c\bar \Lambda_c\pi$ interaction, but there is  a $\Sigma_c\bar \Sigma_c\pi$ interaction. However, because the coupling $g_1$ in Eq.(\ref{baryon}) is  much smaller than $g_2$ and $g_4$ \cite{Lagrangian of Lambda and Sigma}, the interaction  $\Lambda_c\bar \Sigma_c\pi$ is more important. In the $\bar D-\Lambda_c$ system, $\pi^{\pm},\pi^0$ can be exchanged via  the  intermediate state $\Sigma_c$, while there is only one  sort of pion that can be exchanged in the $\bar D-\Sigma_c$ via  the  intermediate state $\Lambda_c$. Therefore, we only consider the $\bar D-\Lambda_c$ system in the following.

The calculation of the two-pion exchange potential is  quite
straightforward. The divergence can be regularized by introducing a gaussian cut-off
 \cite{Ordonez:1995rz} \cite{T.A.Rijken} or in the  dimensional regularization scheme \cite{Kaiser:1997mw}.
Although the scheme of dimensional regularization seems more elegant, it is not suitable for our case. In the dimensional regularization scheme, the  potential  behavior at short distance is $1/r^5$ or $1/r^6$ (which has a singularity  at r=0) \cite{Kaiser:1997mw}, the origin of this  behavior is   the divergent  momentum integral  in the Fourier-transformation to coordinate space. Thus a cutoff is needed  to regularize the short distance behavior;   In the absence of any information about counter terms (short distance interaction), this scheme is also sensitive to the cut-off. The cut-off scheme is widely used for phenomenological estimates. The shortcoming of this scheme is sensitivity to cut-off dependence.
The dependence on the cut-off should be removed by counter terms for which we have no knowledge. But we still can get useful information if we choose a suitable cut-off. For example, a cutoff of the order of  the breakdown scale can
serve as an order-of-magnitude estimate for unknown counter terms.

The calculation is very similar to that for the nucleon-nucleon potential \cite{Ordonez:1995rz} \cite{T.A.Rijken}, except that the parallel box diagram in our case is a little bit different. In the
nucleon case, one should be careful to extract the ``iterated one-pion
exchange'' contribution \cite{Kaiser:1997mw}. Therefore,
``old-fashioned'' time-ordered perturbation theory is widely used
 \cite{Ordonez:1995rz}. However, there is no one-pion exchange
between $D$ and $\Lambda_c$, so we can calculate directly using
covariant perturbation theory.

Mass differences, such as  $a=2(M_{D^*}-M_D$) and
$b=2(M_{\Sigma_c}-M_{\Lambda_c})$,
play  important roles in
our calculation, so we  keep them in the heavy hadron mass expansion. For
instance , we write the propagator of the $D^*$ meson as
\begin{equation}
T_{D^{*}}=-\frac{M_{D^*}g_{\mu\nu}}{(p_{1}-\frac{l-q}{2})^{2}-M_{D^{*}}^{2}+i\epsilon}=-
\frac{g_{\mu\nu}}{-l_{0}-a+i\epsilon},
\end{equation}
where we use the conditions $p_1^2=M_D^2$, the momenta $l$ and $q$ are shown in Fig.1  and $q_0=0$.  Then  the
rest of the calculation can be carried out directly as in
 \cite{Ordonez:1995rz}.

Feynman Diagrams are shown in Fig.1. The corresponding potentials
are given in the Appendix. In  Fig.2, we show the total TPEP with various cut-offs $\Lambda$. At
$\Lambda=1000MeV$, the potential is comparable with that of $DD^*$
at $\Lambda=1200MeV$ in  \cite{Systematicly studied} and
$\Lambda=1000MeV$ in  \cite{I.W.Lee}. Although  the
authors in the latter cases  use the pole-type form factor, we see that there is not much
difference between these two cut-off schemes. For instance,
$\Lambda=1000MeV$ in the Gaussian scheme is comparable with
$\Lambda=900MeV$  in the pole scheme for OPEP  of $DD^*$.
 Uncertainty can   also arise from $g_4$.
 Around the central value of $g_4=1.24$,  there is  only 10 percent uncertainty for $g_4$ which may cause about $100MeV$ uncertainty in  $\Lambda$.

We vary  $\Lambda$ to search the solution of the Schr\"odinger
equation. The result for the bound state of  $\Lambda_c-\bar D$ is shown in the
Table 1. A similar calculation is valid for  the  $\Lambda_b-\bar B$
and the result is shown in  Table.2.



\begin{figure}
\includegraphics[width=18 cm]{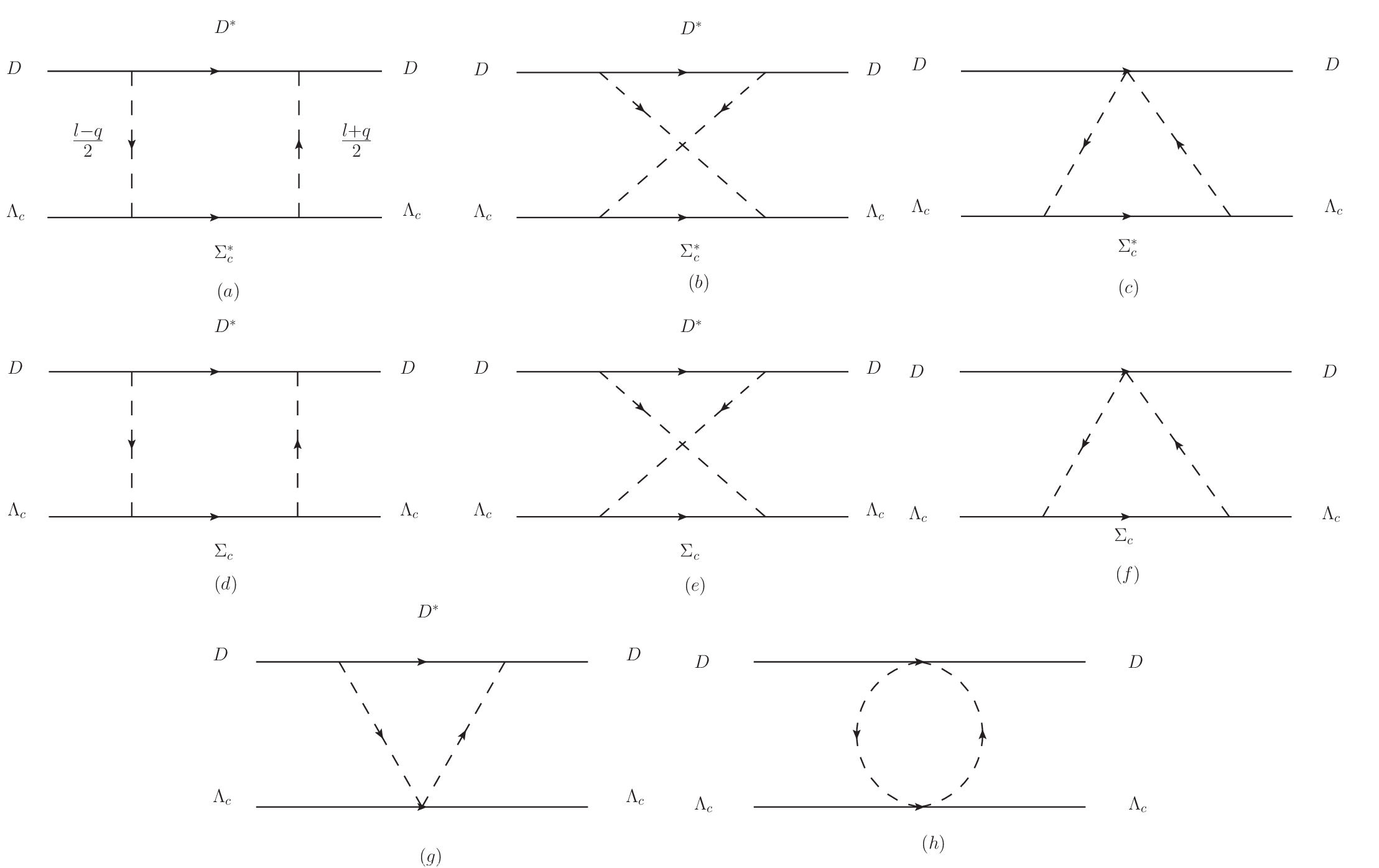}\\
  \caption{The Feynman Diagrams}
\end{figure}

\begin{center}
\begin{figure}
\includegraphics[width=17 cm]{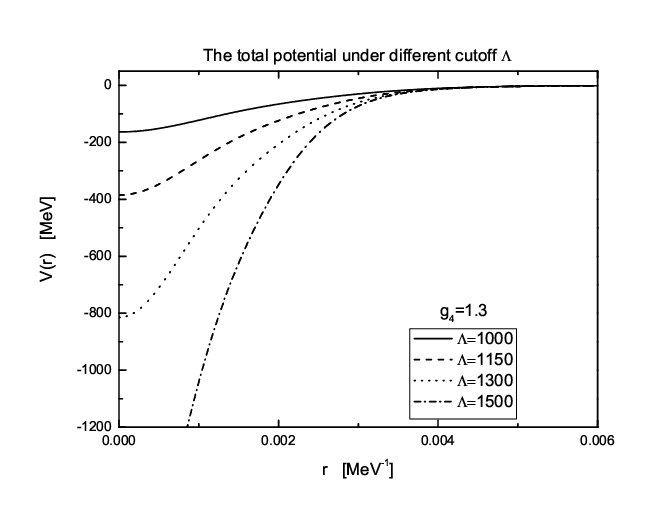}\\
 \caption{The total Potential of the $\Lambda_{c}-\bar D$ with various cutoff $\Lambda$(in units of MeV)}
\end{figure}
\end{center}

\begin{center}
\begin{figure}
\includegraphics[width=17 cm]{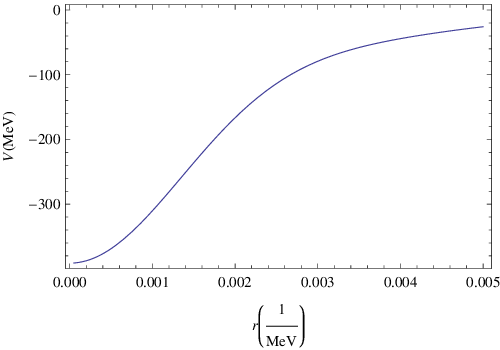}\\
 \caption{TPEP between nucleons with I=1,S=0 at  $\Lambda=700MeV$}
\end{figure}
\end{center}

\begin{center}
\begin{tabular}{|c|c|c|c|}
\hline $\Lambda${[}MeV{]} & $g_{4}$ & $E_{bind}${[}MeV{]} &
$E=M-E_{bind}${[}MeV{]}\tabularnewline \hline \hline 1100 & 1.3 &
Not find & -\tabularnewline \hline 1150 & 1.3 & -1.06 &
4150.20\tabularnewline \hline 1200 & 1.3 & -8.25 &
4143.01\tabularnewline \hline 1250 & 1.3 & -23.74 &
4127.42\tabularnewline \hline
\end{tabular}\\
Table.1 The bound state of the $\bar D-\Lambda_c$ in S-wave.
\par\end{center}

\begin{center}
\begin{tabular}{|c|c|c|c|}
\hline
$\Lambda$ {[}MeV{]} & $g_{4}$ & $E_{bind}$ {[}MeV{]} & $E=M-E_{bind}$ {[}MeV{]}\tabularnewline
\hline
\hline
800 & 1.3 & Not Found & -\tabularnewline
\hline
850 & 1.3 & -0.62 & 10899.08\tabularnewline
\hline
900 & 1.3 & -4.03 & 10895.67\tabularnewline
\hline
950 & 1.3 & -11.58 & 10888.12\tabularnewline
\hline
1000 & 1.3 & -25.05 & 10874.65\tabularnewline
\hline
\end{tabular}\\
Table.2 The bound state of the $\bar B-\Lambda_b$ in S-wave.
\par\end{center}

\section{Discussion and conclusion}
In the cutoff scheme,  the conclusion is inevitably sensitive to  the cutoff. But the value of the cutoff cannot be known a priori.  Because the long distance interaction is cut-off independent,   we would like to know whether the short distance  interaction is overestimated at the cut-off we choose.  We  can address this point by reference  to the deuteron case. By using the TPEP between nucleons obtained in  \cite{T.A.Rijken} plus OPEP, we repeat the process above and find a  $^3S_1(I=0)$  deuteron bound state  starts to appear at $\Lambda=700MeV$
 consistent with  \cite{Systematicly studied}.

 An alternative approach s to study the nucleon's interaction is given in ref. \cite{kolck4}, where the authors use OPEP plus the potential from   the leading order counter terms to fit the experimental data. The potential  from  the counter terms in the leading order of Chiral expansion is written as
\begin{equation}\label{s-d}
 V=C\delta^3(r).
 \end{equation}
In the channel $^3S_1(I=0)$, because the tensor part of OPEP is divergent at $r=0$ the situation is very complicated  \cite{kolck4},  so we make a comparison with the channel $^1S_0$where the value of the coefficient $C$ in (\ref{s-d}) at $\Lambda=700MeV$ is $-(0.74\sim 1)\times 10^{-4}MeV^{-2}$ \cite{kolck4} \cite{kolck3}. Meanwhile, TPEP at $\Lambda=700MeV$ is shown in Fig.3. Analogously, if the potential in Fig.2 at $\Lambda=1000MeV$ is suitable for the heavy meson-baryon system, the  coefficient $C$ of the corresponding potential (\ref{s-d}) could be as large as $-1\times 10^{-4}MeV^{-2}$. Since we do not know $C$ for the heavy meson-baryon system, we refer to the KN system \cite{1405}.
 At $\Lambda=412MeV$, $C=-(2.6\sim 5.7 )\times 10^{-4}MeV^{2}$.

 The dependence of $C$ on $\Lambda$ may be complicated. In  \cite{kolck3},  the authors  use a square well with radius R to smear the delta function of (\ref{s-d}), then $C\sim 1/\Lambda$. This is approximately consistent with the result in  \cite{kolck4}.
 Then,  at $\Lambda=1GeV$, $C=-1\sim -2.3\times 10^{-4}MeV^2$ for the KN system.
 In QCD the interaction between two quarks is mass-independent, so
 the value of $C$ in the KN system may be referred to the $\bar D\Lambda_c$ system.  Surely, $C$  is dependent on the masses of the meson and the baryon. In  \cite{1405}, the value of $C$ increases  with increasing meson mass. If this tendency continues to the mass of the $D$ meson, $\Lambda$  is in the range  $1000MeV\sim 1150MeV$.

From the above  consideration, a bound state of  the $B-\Lambda_b$ is  optimistically expected, while for the $\bar D-\Lambda_c$ case the bound state is prudently expected  and the situation may be dependent on the higher order corrections  of the heavy quark expansion. Referring to the $X(3872)$  which is considered as  a bound state of $D-\bar D^*$ as many authors
suggested \cite{3872}, if the $\bar D-\Lambda_c$ bound state
($\Lambda(4150)_{cc}$)  exists,
it  could be produced in the channel $\Lambda_b\to\Lambda(4150)_{cc} K\to\eta_c N K$. The exact branching ratio of this channel is not easy to obtain, but  the order of the branching ratio  $\Lambda_b\to\Lambda(4150)_{cc} K$ may
be roughly equal to that of $B\to X(3872) K$, because the scalar diquark in $\Lambda_b$ and the light quark in $B$ could be roughly considered as  spin-decoupled spectators in heavy quark limit. This channel is expected to be seen in LHCb.

\section{Acknowledgements}
This work is supported partly by NNSFC under grant 11175153/A050202
and the Fundamental Research Funds for the Central Universities. We
would like to thank Prof. Jifeng Yang for  very helpful discussions.
\section{Appendix}

1. spin-$\frac{3}{2}$ digram in momentum space

\begin{eqnarray*}
V_a(\vec{q}) & = & \frac{-1}{4}(\frac{g}{f_{\pi}})^{2}(\frac{g_{4}}{f_{\pi}})^{2}\intop\frac{d^{3}l}{(2\pi)^{3}}(\vec{l}-\vec{q})\cdot(\vec{l}+\vec{q})(\vec{l}-\vec{q})\cdot(\vec{l}+\vec{q})\{\frac{1}{2(a+c)}[\frac{1}{\omega_{1}(\omega_{1}+a)}-\frac{1}{\omega_{2}(\omega_{2}+a)}]\frac{1}{\omega_{2}^{2}-\omega_{1}^{2}}\\
 &  & +\frac{1}{2(a+c)}[\frac{1}{\omega_{1}(\omega_{1}+c)}-\frac{1}{\omega_{2}(\omega_{2}+c)}]\frac{1}{\omega_{2}^{2}-\omega_{1}^{2}}\},\\
V_b(\vec{q}) & = & \frac{-1}{4}(\frac{g}{f_{\pi}})^{2}(\frac{g_{4}}{f_{\pi}})^{2}\intop\frac{d^{3}l}{(2\pi)^{3}}(\vec{l}-\vec{q})\cdot(\vec{l}+\vec{q})(\vec{l}-\vec{q})\cdot(\vec{l}+\vec{q})\{\frac{-1}{2(a-c)}[\frac{1}{\omega_{1}(\omega_{1}+a)}-\frac{1}{\omega_{2}(\omega_{2}+a)}]\frac{1}{\omega_{2}^{2}-\omega_{1}^{2}}\\
 &  & +\frac{1}{2(a-c)}[\frac{1}{\omega_{1}(\omega_{1}+c)}-\frac{1}{\omega_{2}(\omega_{2}+c)}]\frac{1}{\omega_{2}^{2}-\omega_{1}^{2}}\},\\
V_c(\vec{q}) & = & -\frac{1}{2}(\frac{g_{4}}{f_{\pi}})^{2}\frac{1}{f_{\pi}^{2}}\intop\frac{d^{3}l}{(2\pi)^{3}}(\vec{l}-\vec{q})\cdot(\vec{l}+\vec{q})(\frac{1}{\omega_{1}+c}-\frac{1}{\omega_{2}+c})\frac{1}{\omega_{2}^{2}-\omega_{1}^{2}},\end{eqnarray*}

where
\[
\omega_{1}=\sqrt{(\vec{q}-\vec{l})^{2}+4m_{\pi}^{2}},\]

\[
\omega_{2}=\sqrt{(\vec{q}+\vec{l})^{2}+4m_{\pi}^{2}},\]

and
\begin{center}
$a=2(M_{D^{*}}-M_{D})$, $b=2(M_{\Sigma_{c}}-M_{\Lambda_{c}})$,
$c=2(M_{\Sigma_{c}^{*}}-M_{\Lambda_{c}})$.
\par\end{center}

we transform above into coordinate space and obtain:

\begin{eqnarray*}
V_a(r) & = & \frac{-1}{4\pi}(\frac{g}{f_{\pi}})^{2}(\frac{g_{4}}{f_{\pi}})^{2}\frac{1}{a+c}\int d\lambda\frac{a}{a^{2}+\lambda^{2}}\left[\frac{2}{r^{2}}F^{'}\left(\lambda,r\right)F^{'}\left(\lambda,r\right)+F^{''}\left(\lambda,r\right)F^{''}\left(\lambda,r\right)\right]\\
 &  & +\frac{c}{c^{2}+\lambda^{2}}\left[\frac{2}{r^{2}}F^{'}\left(\lambda,r\right)F^{'}\left(\lambda,r\right)+F^{''}\left(\lambda,r\right)F^{''}\left(\lambda,r\right)\right],\\
V_b(r) & = & \frac{-1}{4\pi}(\frac{g}{f_{\pi}})^{2}(\frac{g_{4}}{f_{\pi}})^{2}\frac{1}{a-c}\int d\lambda\frac{-a}{a^{2}+\lambda^{2}}\left[\frac{2}{r^{2}}F^{'}\left(\lambda,r\right)F^{'}\left(\lambda,r\right)+F^{''}\left(\lambda,r\right)F^{''}\left(\lambda,r\right)\right]\\
 &  & +\frac{c}{c^{2}+\lambda^{2}}\left[\frac{2}{r^{2}}F^{'}\left(\lambda,r\right)F^{'}\left(\lambda,r\right)+F^{''}\left(\lambda,r\right)F^{''}\left(\lambda,r\right)\right],\\
V_c(r) & = & -\frac{1}{\pi}(\frac{g_{4}}{f_{\pi}})^{2}\frac{1}{f_{\pi}^{2}}\int d\lambda\frac{\lambda^{2}}{c^{2}+\lambda^{2}}\left(F^{'}\left(\lambda,r\right)F^{'}\left(\lambda,r\right)\right),\end{eqnarray*}

2.spin-$\frac{1}{2}$ in momentum space\begin{eqnarray*}
V_d(\vec{q}) & = & \frac{-3}{8}(\frac{g}{f_{\pi}})^{2}(\frac{g_{2}}{f_{\pi}})^{2}\intop\frac{d^{3}l}{(2\pi)^{3}}(\vec{l}-\vec{q})\cdot(\vec{l}+\vec{q})(\vec{l}-\vec{q})\cdot(\vec{l}+\vec{q})\{\frac{1}{2(a+b)}[\frac{1}{\omega_{1}(\omega_{1}+a)}-\frac{1}{\omega_{2}(\omega_{2}+a)}]\frac{1}{\omega_{2}^{2}-\omega_{1}^{2}}\\
 &  & +\frac{1}{2(a+b)}[\frac{1}{\omega_{1}(\omega_{1}+b)}-\frac{1}{\omega_{2}(\omega_{2}+b)}]\frac{1}{\omega_{2}^{2}-\omega_{1}^{2}}\},\\
V_e(\vec{q}) & = & \frac{-3}{8}(\frac{g}{f_{\pi}})^{2}(\frac{g_{2}}{f_{\pi}})^{2}\intop\frac{d^{3}l}{(2\pi)^{3}}(\vec{l}-\vec{q})\cdot(\vec{l}+\vec{q})(\vec{l}-\vec{q})\cdot(\vec{l}+\vec{q})\{\frac{-1}{2(a-b)}[\frac{1}{\omega_{1}(\omega_{1}+a)}-\frac{1}{\omega_{2}(\omega_{2}+a)}]\frac{1}{\omega_{2}^{2}-\omega_{1}^{2}}\\
 &  & +\frac{1}{2(a-b)}[\frac{1}{\omega_{1}(\omega_{1}+b)}-\frac{1}{\omega_{2}(\omega_{2}+b)}]\frac{1}{\omega_{2}^{2}-\omega_{1}^{2}}\},\\
V_f(\vec{q}) & = & -\frac{3}{4}(\frac{g_{2}}{f_{\pi}})^{2}\frac{1}{f_{\pi}^{2}}\intop\frac{d^{3}l}{(2\pi)^{3}}(\vec{l}-\vec{q})\cdot(\vec{l}+\vec{q})(\frac{1}{\omega_{1}+b}-\frac{1}{\omega_{2}+b})\frac{1}{\omega_{2}^{2}-\omega_{1}^{2}},\end{eqnarray*}

and in  coordinate space

\begin{eqnarray*}
V_d(r) & = & \frac{-3}{8\pi}(\frac{g}{f_{\pi}})^{2}(\frac{g_{2}}{f_{\pi}})^{2}\frac{1}{a+c}\int d\lambda\frac{a}{a^{2}+\lambda^{2}}\left[\frac{2}{r^{2}}F^{'}\left(\lambda,r\right)F^{'}\left(\lambda,r\right)+F^{''}\left(\lambda,r\right)F^{''}\left(\lambda,r\right)\right]\\
 &  & +\frac{b}{b^{2}+\lambda^{2}}\left[\frac{2}{r^{2}}F^{'}\left(\lambda,r\right)F^{'}\left(\lambda,r\right)+F^{''}\left(\lambda,r\right)F^{''}\left(\lambda,r\right)\right],\\
V_e(r) & = & \frac{-3}{8\pi}(\frac{g}{f_{\pi}})^{2}(\frac{g_{2}}{f_{\pi}})^{2}\frac{1}{a-c}\int d\lambda\frac{-a}{a^{2}+\lambda^{2}}\left[\frac{2}{r^{2}}F^{'}\left(\lambda,r\right)F^{'}\left(\lambda,r\right)+F^{''}\left(\lambda,r\right)F^{''}\left(\lambda,r\right)\right]\\
 &  & +\frac{b}{b^{2}+\lambda^{2}}\left[\frac{2}{r^{2}}F^{'}\left(\lambda,r\right)F^{'}\left(\lambda,r\right)+F^{''}\left(\lambda,r\right)F^{''}\left(\lambda,r\right)\right],\\
V_f(r) & = & -\frac{3}{2\pi}(\frac{g_{2}}{f_{\pi}})^{2}\frac{1}{f_{\pi}^{2}}\int d\lambda\frac{\lambda^{2}}{b^{2}+\lambda^{2}}\left(F^{'}\left(\lambda,r\right)F^{'}\left(\lambda,r\right)\right),\end{eqnarray*}

There is a triangle diagram(digram g in Fig.1) and a 4-veterx digram(digram h in Fig.1) without $\Sigma$ or $\Sigma^{*}$ propagator which provides :

\begin{eqnarray*}
V_g(\vec{q}) & = & -\frac{3}{4}(\frac{g}{f_{\pi}})^{2}\frac{1}{f_{\pi}^{2}}\intop\frac{d^{3}l}{(2\pi)^{3}}(\vec{l}-\vec{q})\cdot(\vec{l}+\vec{q})(\frac{1}{\omega_{1}+a}-\frac{1}{\omega_{2}+a})\frac{1}{\omega_{2}^{2}-\omega_{1}^{2}},\\
V_h(\vec{q}) & = & -\frac{3}{8}\frac{1}{f_{\pi}^{4}}\intop\frac{d^{3}l}{(2\pi)^{3}}\frac{1}{\omega_{1}+\omega_{2}},\\
V_g
(r) & = & -\frac{3}{2\pi}(\frac{g}{f_{\pi}})^{2}\frac{1}{f_{\pi}^{2}}\int d\lambda\frac{\lambda^{2}}{a^{2}+\lambda^{2}}\left(F^{'}\left(\lambda,r\right)F^{'}\left(\lambda,r\right)\right),\\
V_h(r) & = & -\frac{3}{4\pi}\frac{1}{f_{\pi}^{4}}\int d\lambda\lambda^{2}F^{2}(\lambda,r).\end{eqnarray*}

The total potential of the $D-\Lambda_{c}$  is :

\begin{eqnarray*}
V(r) & = & V_a(r)+V_b(r)+V_c(r)+V_d(r)\\
 &  & +V_e(r)+V_f(r)+V_g(r)+V_h(r).\end{eqnarray*}

 The Function $F(\lambda,r)$ used in the potential is defined as
\noindent \begin{eqnarray*}
F(\lambda,r) & = & e^{-\frac{\lambda^{2}}{\Lambda^{2}}}I_{2}(\sqrt{(2m_{\pi})^{2}+\lambda^{2}},r),\\
I_{2}(m,r) & = & \frac{m}{4\pi}\phi_{c}^{0}(m,r),\\
F^\prime(\lambda,r) &=& \displaystyle{\frac{d}{dr}F(\lambda,r)},\\
\phi_{c}^{0}(m,r) & = & e^{\frac{m^{2}}{\Lambda^{2}}}\left[e^{-mr}Erfc(-\frac{\Lambda r}{2}+\frac{m}{\Lambda})-e^{mr}Erfc(\frac{\Lambda r}{2}+\frac{m}{\Lambda})\right]\frac{1}{2mr}.
\end{eqnarray*}

Here the function $Erfc(x)$ is the complementary error function. More
details can be found in the Appendix of Ref \cite{T.A.Rijken}.

\end{document}